\documentclass[%
 aip,
cp,  
 amsmath,amssymb,
 reprint,%
]{revtex4-2}

\usepackage{graphicx}
\usepackage{dcolumn}
\usepackage{bm}

\usepackage[utf8]{inputenc}
\usepackage[T1]{fontenc}
\usepackage{mathptmx}

\def\la{\langle}
\def\ra{\rangle}

\def\be{\begin{equation}}
	\def\ee{\end{equation}}
\def\la{\langle}
\def\ra{\rangle}

\newcommand{\beq}{\begin{eqnarray}}
	\newcommand{\eeq}{\end{eqnarray}}
\def\be{\begin{equation}}
	\def\ee{\end{equation}}
\newcommand\eq[1]{Eq.~(\ref{#1})}

\def\dd{\partial}

\begin{document}

\title{Casimir Force, Magnetization, and Susceptibility for \\
	Ising Chain with an Impurity Atom in the Presence of an External Field}

\author{Daniel Dantchev} 
 \email[Corresponding author: ]{danieldantchev@gmail.comu}
\affiliation{
 Institute of
 Mechanics, Bulgarian Academy of Sciences, Academic Georgy Bonchev St. building 4,
 1113 Sofia, Bulgaria.
}


\date{\today} 

\begin{abstract}
The article contains exact results for the behavior of the Casimir force of Ising chain of $N$ atoms containing an impurity atom. The force is considered as a function of $N$ and on both the temperature $T$ and the external magnetic field  $h$. The interactions between nearest neighbors within the chain are supposed to be of a ferromagnetic short-ranged type $J>0$. The impurity atom interacts with constants $J_{ia}$ with its nearest neighbors, where $J_{ia}$ can be both of ferromagnetic type, i.e., $J_{ia}>0$, as well as of antiferromagnetic type, i.e., $J_{ia}<0$.  It turns out that the Casimir force can be both attractive or repulsive, depending on the sign and the magnitude of the ratio $J_{ia}/J$. 
\end{abstract}

\maketitle

\section{\label{sec:level1}Introduction}

In the general case, we envisage a  $d$-dimensional system with a film geometry $\infty^{d-1}\times L$, $L\equiv L_\perp$, and with boundary conditions $\zeta$ imposed along the spatial direction of finite extent $L$.  Let us take ${\cal F}_{ {\rm tot}}^{(\zeta)}(L,T,h)$ to be the total (Gibbs) free energy of such a system within the GCE, where $T$ is the temperature and $h$ is the magnetic field. Then, if   $f^{(\zeta)}(T,h,L)\equiv \lim_{A\to\infty}{\cal F}_{ {\rm tot}}^{(\zeta)}/A$  is the free energy per area $A$ of the system, one can define the Casimir force for critical systems in the grand-canonical $(T-h)$-ensemble, see, e.g. Ref. \cite{Krech1994,BDT2000,MD2018,DD2022,Gambassi2024}, as: 
\begin{equation}
	\label{CasDef}
	\beta F_{\rm Cas}^{(\zeta)}(L,T,h)\equiv- \frac{\partial}{\partial L}f_{\rm ex}^{(\zeta)}(L,T,h)
\end{equation}
where
\begin{equation}
	\label{excess_free_energy_definition}
	f_{\rm ex}^{(\zeta)}(L,T,h) \equiv f^{(\zeta)}(L,T,h)-L f_b(T,h)
\end{equation}
is the so-called excess (over the bulk) free energy per area and per $\beta^{-1}=k_B T$.

	\section{\label{sec:the model} The Model}
\sectionmark{Ising Model}

Let us consider a one-dimensional Ising chain of $N$ spins $(S_i \pm 1, i=1,.., N)$. We suppose that the interaction between them $J$ is of a ferromagnetic type, i.e., that $J>0$.  The Hamiltonian of the model is given by
\begin{equation}
	\label{eq:Hamiltonian}
	{\cal	H}=-J\sum_{i=2}^{N-1} S_i S_{i+1} - J_{ia}\; S_1(S_2+S_1)+ \hat{h}\sum_{i=1}^N S_i.
\end{equation}
Obviously, in the special case $J_{ia}=J$ the systems becomes equivalent to one under periodic boundary conditions, for which case the results are well known  see, e.g., \cite{MW73,H87,B82,Kardar2007,Pathria2011} for the behavior of the infinite chain and \cite{Rudnick2010,DD2022,Dantchev2023} for the finite one. 

\begin{figure}[h!]
	\centering
	\includegraphics[width=\columnwidth]{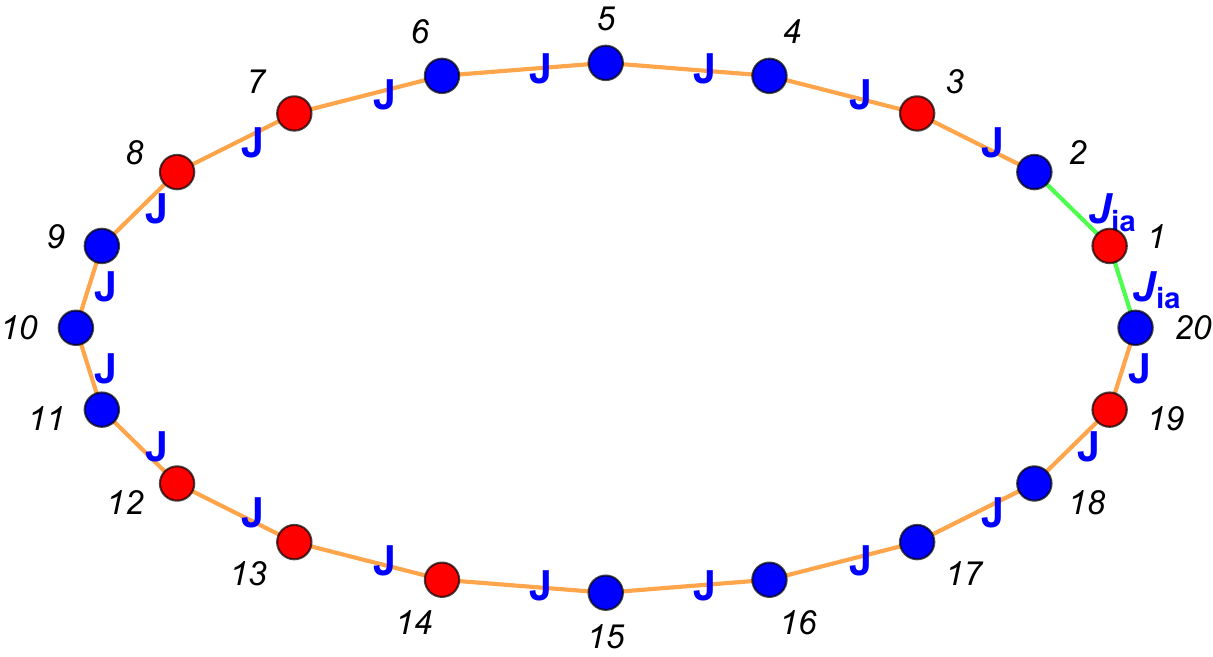} \quad
	\caption{Illustration of the definition of the model. The figure shows a set of $N=20$ spins (circles) with two possible values $+1$ (say red) and $-1$ (blue). There are interaction bonds between any two spins equal either to $J$, or to $J_{ia}$. In the depicted example we have 8 spins $+1$, and 12 spins $-1$. The last implies that the total magnetization of the chain is $M=\sum_{i=1}^{20} S_i=-4$.}
	\label{fig:the-model}
\end{figure}

A spin configuration is denoted by ${\bf S}=\left\{S_1, S_2, \cdots, S_N \right\} \in \left\{-1,1\right\}^N$. For any $\beta\in {\mathbb R}^+, h\in {\mathbb{R}}$ and $N\in \mathbb{N}^+$ the Ising model is defined by the Boltzmann distributions
\begin{equation}
	\label{eq:Boltzmann-distribution}
	\mu_N({\bf S}|\beta,h,J,J_{ia})=\frac{1}{Z(N,K,K_{ia}, h)}\exp\left[-\beta  {\cal H}\right],
\end{equation}
where the normalization $Z(N,K,K_{ia}, h)$ is known as the partition function.
This function is one of the main quantities of interest in the statistical mechanics. It connects the mathematics to the physical potential in the thermodynamics, usually called free energy  $f(N,K,K_{ia},h)$ 
\begin{equation}
	\label{eq:general-free-energy}
	\beta f(N,K,K_{ia},h) = -\frac{1}{N} \ln Z(N,K,K_{ia},h). 
\end{equation}
For the considered model the partition function  is given by
\begin{equation}
	Z(N,K,K_{ia}, h)=\sum_{\{S_i\}} \exp\left[-\beta  {\cal H}\right].
\end{equation}
Here $K=\beta J$, $K_{ia}=\beta J_{ia}$ and $h=\beta \hat{h}$. 
The definition of the model is illustrated in Fig. \ref{fig:the-model}. 

For any function $f: \{-1, +1\}^N \to {\mathbb R}$ one  defines the average
$\la f({\bf S}) \ra_{\mu_N}$ as 
\begin{equation}
	\label{eq:average-definition}
	\la f({\bf S}) \ra_{\mu_N} = \sum_{{\bf S}\in {\{-1,+1\}^N}} f({\bf S})\; \mu_N({\bf S}). 
\end{equation}

\section{On the Behavior of the free energy of the Ising chain  with an impurity atom}
\label{sec:Ising-free-energy}

The partition function of the system can be written in the form 
\begin{eqnarray}
	\label{ZIsingH}
	Z(N,K,K_{ia}, h) &=& \sum_{\{S_i=\pm 1\}} \exp \left[K\left(S_2 S_3+\cdots + S_{N-2} S_{N-1} \right) +K_{ia} S_1 (S_2+S_N)\right. \nonumber \\
	&&\left.+h\left(S_1+S_2+\cdots+S_{N-1}+ S_N\right)\right]. 
\end{eqnarray}
It is helpful to cast the above formula as follows 
\begin{eqnarray}
	\label{ZIsingHeq}
	&&	Z(N,K,K_{ia}, h)= \\
	&& \sum_{\{S_i\}}\exp \left[\frac{1}{2} \left(h S_1+h S_2\right)+K_{ia} S_1 S_2\right] \exp \left[\frac{1}{2} \left(h S_2+h S_3\right)+K S_2 S_3\right]\nonumber
	\\
	&& \times \cdots \times \exp \left[\frac{1}{2} \left(h S_{N-1}+h S_N\right)+K S_{N-1} S_N\right] \exp \left[\frac{1}{2} \left(h S_{N}+h S_1\right)+K_{ia} S_{N} S_1\right]. \nonumber
\end{eqnarray}
Introducing the matrices 
\begin{equation}
	\label{eq:matrices_internal_part_chain}
	{\mathbf T}=\begin{pmatrix}
		\exp\left(K+h\right) & \exp\left(-K\right) \\\\
		\exp\left(-K\right) & \exp\left(K-h \right)    
	\end{pmatrix}, \quad \mbox{and} \quad 	{\mathbf T}_{ia}=\begin{pmatrix}
		\exp\left(K_{ia}+h\right) & \exp\left(-K_{ia}\right) \\\\
		\exp\left(-K_{ia}\right) & \exp\left(K_{ia}-h \right)    
	\end{pmatrix},
\end{equation}
it is easy to show that 
\begin{equation}
	\label{eq:Z-via_T}
	Z(N,K,K_{ia}, h)=	{\rm Tr}\,\left[{\mathbf T}_{ia}{\mathbf T}^{N-1}  {\mathbf T}_{ia}\right].
\end{equation}
The matrix ${\mathbf T}$ is usually called transfer matrix.  In the remainder we will need its eigenvalues. They are, customarily, denoted by  $\lambda_{1}$ and $\lambda_2$ and read 
\be
\label{eq:eigenvalues_per1}
\lambda_{1,2}(K, h) = e^K \left(\cosh h \pm \sqrt{e^{-4 K} + \sinh^2 h}\right).
\ee
Obviously, $\lambda_{1},\lambda_2 \in \mathbb{R}$, with $\lambda_{1}>\lambda_2$. The two-dimensional  matrix, which diagonalizes ${\mathbf T}$ is 
\begin{equation}
	\label{eq:orthogonal_matrix_P}
	\mathbf{P}=\begin{pmatrix}
		\cos\phi & -\sin \phi \\ \\
		\sin \phi & \cos \phi  
	\end{pmatrix}, \quad \mbox{with} \quad \mathbf{P}^{-1}\,\mathbf{T}\, \mathbf{P} =\begin{pmatrix}
		\lambda_{1} & 0 \\ \\
		0 & \lambda_2 
	\end{pmatrix},
\end{equation}
where $\phi$ is determined by \cite{B82}
\begin{equation}
	\label{eq:def-phi}
	\cot 2\phi=\exp(2K)\sinh h, \quad 0<\phi<\pi/2.
\end{equation}
Explicitly, one has
\begin{eqnarray}
	\label{eq:matrix-P-explicit}
	\cos \phi &=& \frac{1}{\sqrt{2}}\sqrt{1+\frac{\sinh (h)}{\sqrt{\sinh ^2(h)+e^{-4 K}}}}\,, \nonumber\\
	\sin \phi &=& \frac{1}{\sqrt{2}}\sqrt{1-\frac{\sinh (h)}{\sqrt{\sinh ^2(h)+e^{-4 K}}}}\,\\
	\sin (2 \phi ) &=& \frac{1}{\sqrt{\cot ^2(2 \phi )+1}}=\frac{1}{\sqrt{\left[e^{2 K} \sinh (h)\right]^2+1}}.\nonumber
\end{eqnarray}
Using the cyclic property of the trace operation we can transform $Z(N,K,K_a, h)$ in \eq{eq:Z-via_T} into
\begin{eqnarray}
	\label{eq:diagonalization}
	Z(N,K,K_{ia}, h) &=& {\rm Tr}\,\left[\left(\mathbf{P}^{-1} {\mathbf T}^{N-2}\mathbf{P}\right)\, \left(\mathbf{P}^{-1} {\mathbf T}_{ia}^2 \, \mathbf{P}\right)\right]\nonumber\\
	&=&{\rm Tr}\,\left[\begin{pmatrix}
		\lambda_{1}^{N-1} & 0 \\ \\
		0 & \lambda_2^{N-1} 
	\end{pmatrix}\right]\left(\mathbf{P}^{-1} {\mathbf T}_{ia}^2 \, \mathbf{P}\right).
\end{eqnarray}
Performing the calculations in \eq{eq:diagonalization}, we obtain the partition function of the finite Ising chain with an impurity atom:
\begin{eqnarray}
	\label{eq:Z-final-result}
	&&	Z(N,K,K_{ia}, h) =	\lambda _1^{N-2} \left[e^{-2 (h+ K_{ia})} \left(e^{2 K_{ia}} \left(e^{2 h} \cos ^2(\phi )+\sin ^2(\phi )\right)+e^h \sin (2 \phi )\right)^2\right]  \nonumber\\
	&& +\lambda _2^{N-2} \left[e^{-2 (h+K_{ia})} \left(e^{2 (h+K_{ia})} \sin ^2(\phi )- e^h \sin (2\phi ) +e^{2 K_{ia}} \cos ^2(\phi )\right)^2\right]. 
\end{eqnarray}
Using \eq{eq:matrix-P-explicit}, the above can be rewritten in explicit form
\begin{equation}
	\label{eq:Z-eq}
	Z(N,K,K_{ia}, h) =	\lambda _1^{N-2} a_1(K,K_{ia}, h) +	\lambda _2^{N-2} a_2(K,K_{ia}, h),
\end{equation}
where 
\begin{eqnarray}
	\label{eq:a1-and-a2}
	a_1(K,K_{ia}, h)&=& \frac{e^{-2 K_{ia}} \left(1+\sinh ^2(h) e^{2 (K+K_{ia})}+e^{2 K_{ia}} \cosh (h) \sqrt{e^{4 K} \sinh ^2(h)+1}\right)^2}{e^{4 K} \sinh ^2(h)+1} \;\; \; \;\\
	a_2(K,K_{ia}, h)&=& \frac{e^{-2 K_{ia}} \left(1+\sinh ^2(h) e^{2 (K+K_{ia})}-e^{2 K_{ia}} \cosh (h) \sqrt{e^{4 K} \sinh ^2(h)+1}\right)^2}{e^{4 K} \sinh ^2(h)}. \nonumber
\end{eqnarray}

\subsection{On the Behavior of the Infinite Ising Chain }
\label{sec:bulk-system}

The behavior in GCE of the one-dimensional 
Ising model in the thermodynamic limit $N\to \infty$  is discussed in most textbooks on statistical 
mechanics \cite{B82,H87,Kardar2007,Pathria2011,Liebchen2019} and is well known.  Here we summarize only these of the known results that will be used in the remainder of the current study. 

We first remind that the infinite one-dimensional Ising chain with short-ranged interactions exhibits an essential critical point at $T=0$. About that point the chain demonstrates the usual scaling behavior - see below.  

The free energy density of the "bulk" chain is 
\begin{equation}
	\label{eq:bulk-gibbs-free-energy}
	\beta f_b(K,h)=-\ln \lambda_{1}(K,h),
\end{equation}
with $ \lambda_{1}(K,h)$ given in \eq{eq:eigenvalues_per1}. 
We know the behavior of the correlation length, see, e.g.,  Ref. \cite[p. 36, Eq. 2.2.15]{B82} 
\begin{equation}
	\label{eq:scaling_length_Ising}
	\xi^{-1}(K,h)=\ln \left[\lambda_1(K,h)/\lambda_2(K,h)\right],
\end{equation}
and so one can easily specify the scaling variables.  It is clear that $\xi$ diverges when $\lambda_2\to\lambda_{1}$. Obviously, this happens when $h\to 0$ and $K\to\infty$.  Defining
\be
\label{eq:xit_Ising}
\xi_t\equiv\xi(K,0)\simeq \frac{1}{2}e^{2K}, \;\mbox{when}\; K\gg 1, \mbox{and}\; \xi_h\equiv\lim_{K\to \infty}\xi(K,h)\simeq \frac{1}{2h}, \; \mbox{when} \; h\ll 1,
\ee
for the scaling variables one identifies
\be\label{eq:scaling_variables_Ising}
x_t=N/\xi_t= 2 N e^{-2K}, \quad \mbox{and} \quad x_h =N/\xi_h=2 N h. 
\ee
It immediately follows that the correlation length, the bulk magnetization, and the bulk Gibbs free energy in the limit $K\gg 1$,  in terms of these scaling variables,  read 
\be
\label{eq:corr_length_scaling}
\xi(K,h)=\frac{N}{\sqrt{x_h^2+x_t^2}}, \quad m_b(K,h)=\frac{x_h}{\sqrt{x_h^2+x_t^2}}, \quad \beta f_b(K,h)= -K-\frac{1}{4 N}\sqrt{x_h^2+x_t^2}.
\ee

Next, it must be recalled that in terms of $t=\exp(-2K)$ and $h$, one obtains the usual scaling relations with the scaling exponents, see, e.g., Ref. \cite{B82} 
\be
\label{eq:crit_exponents}
\alpha=\gamma=\nu=\eta=1,\; \beta=0,\;\delta=\infty, \; \mbox{with, however,}\; \beta \delta=1. 
\ee

\subsection{On the leading  finite-size corrections for the finite chain with impurity atom}

By definition, the free energy density of the finite chain is,
\begin{eqnarray}
	\label{eq:chain-gibbs-free-energy}
	\beta f(N,K,K_{ia},h)&=& -\frac{1}{N} \ln Z(N,K,K_{ia},h)\\
	&=& -\ln \lambda_{1}(K,h) +\ln\left[\lambda_1^2(K,h)-	a_1(K,K_{ia}, h)\right]\nonumber\\
	&& -\frac{1}{N}\ln \left\{1+\exp \left[-\frac{N-2}{\xi (K,h)}\right] \;\frac{a_2(K,K_{ia}, h)}{a_1(K,K_{ia}, h)}\right\}. \nonumber
\end{eqnarray}
From \eq{eq:chain-gibbs-free-energy} and 	\eq{eq:Z-final-result}, one obtains the result 
\begin{equation}
	\label{eq:decomposition-of-free-energy}
	\beta f(N,K,K_{ia},h) = \beta f_b(K,h)+\frac{1}{N}\beta f_{\rm surface}(K,K_{ia},h)+ \beta \Delta f(N,K,K_{ia},h),
\end{equation}
where
\begin{equation}
	\label{eq:surface-free-energy}
	\beta f_{\rm surface}(K,K_a,h)= \ln\left[\lambda_1^2(K,h)-	a_1(K,K_{ia}, h)\right],
\end{equation}
and 
\begin{eqnarray}
	\label{eq:remnant-part}
	\beta \Delta f(N,K,K_{ia},h)&=&-\frac{1}{N}\ln \left\{1+\exp \left[-\frac{N-2}{\xi (K,h)}\right] \;	r(K,K_{ia},h)\right\}. 
\end{eqnarray}
Here 
\begin{eqnarray}
	\label{eq:ratio-of-a1-and-a2}
	r(K,K_{ia},h)&=&\frac{a_2(K,K_{ia}, h)}{a_1(K,K_{ia}, h)}=\left(\frac{\frac{e^{2 K} \sinh ^2(h)+e^{-2 K_{ia}}}{\cosh (h) \sqrt{e^{4 K} \sinh ^2(h)+1}}-1}{\frac{e^{2 K} \sinh ^2(h)+e^{-2 K_{ia}}}{\cosh (h) \sqrt{e^{4 K} \sinh ^2(h)+1}}+1}\right)^2 \nonumber\\
	&=&\left(1-\frac{2}{\frac{e^{2 K} \sinh ^2(h)+e^{-2 K_{ia}}}{\cosh (h) \sqrt{e^{4 K} \sinh ^2(h)+1}}+1}\right)^2>0.
\end{eqnarray}

Let us briefly discuss the meaning of the terms in 	\eq{eq:decomposition-of-free-energy}.

First, the term $\beta f_{\rm surface}(K,K_{ia},h)$ has the  meaning of a "surface free energy density". It shall exist in s system with an impurity atom and is due to the contribution of the interaction of this atom with neighbor that differs from those of the interactions of other atoms. It is, however clear this term shall vanish for PBC's. One can check that, after some simple algebra, indeed $\lambda_1^2(K,h)-	a_1(K,K_{ia}=K, h)=0$. 

Second, we note that the redundant term $\beta \Delta f(N,K,K_{ia},h)$ is exponentially small for $N\gg 1$ when $\xi (K,h)={\cal O}(1)$. The only  exception is the case when $N\propto \xi (K,h)$, i.e., $N/\xi (K,h)={\cal O}(1)$. The last relation defines the so-called "finite-size scaling region".  Inherently, it is described by scaling variables $x_t={\cal O}(1)$ and $x_h={\cal O}(1)$, as given in \eq{eq:scaling_variables_Ising}. Thus, the term $\beta \Delta f(N, K,K_{ia},h)$ cannot  be neglected and, as we will see below, is of a primarily importance for the behavior of the Casimir force.

\section{On the behavior of the magnetization}

One of the functions of the random variables $\{S_i\}, i=1,\cdots,N$ which has a well known physical meaning is the average of $1/N\sum_{i=1}^N S_i$. More specifically, $m(N,K,K_{ia},h)=\la 1/N\sum_{i=1}^N S_i \ra_{\mu_N}$ is called average magnetization. It is easy to check that
\begin{equation}
	\label{eq:average-magnetization}
	m(N,K,K_{ia},h)=-\frac{\dd}{\dd h} \left[\beta f(N,K,K_{ia},h)\right]. 
\end{equation}
Having into account the symmetry of the system under the transformation $S-i\to -S_i$ for all spins, together with $h \to -h$, it is clear that 
\begin{equation}
	\label{eq:symmetry-properties}
	f(N,K,K_{ia},h)=f(N,K,K_{ia},-h) \quad \mbox{and} \quad m(N,K,K_{ia},h)=-m(N,K,K_{ia},-h),
\end{equation}
i.e., $f(N,K,K_{ia},h)$ is an even, while $m(N,K,K_{ia},h)$ is and odd function of $h$. 
\begin{figure}[tbh!]
	\centering
	\includegraphics[width=\columnwidth]{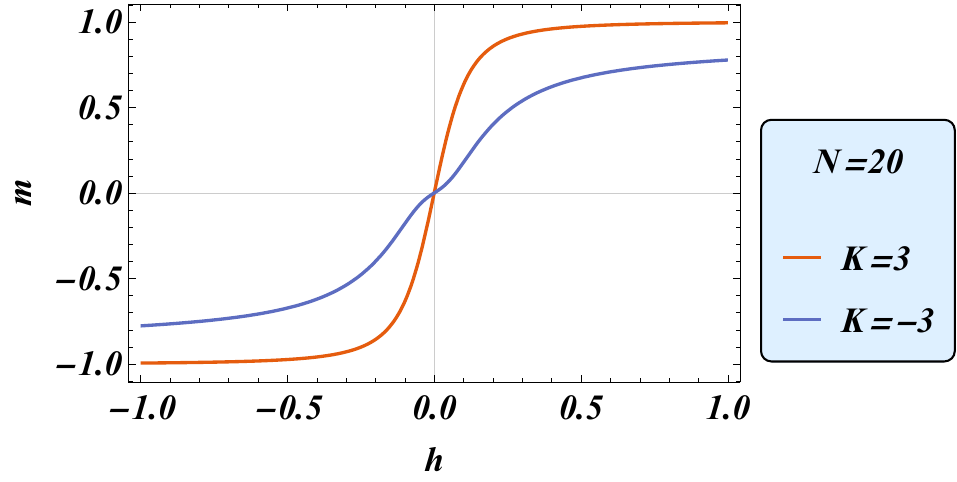} \quad
	\caption{The behavior of the average magnetization for $K=3$ and $K=-3$ for a chain with $N=20$ random (dynamical) variables.}
	\label{fig:magnetization}
\end{figure}
The behavior of the magnetization as a function of $h$ for $N=20$, $K=1$, $K_{ia}=\pm 3$ is visualized in Fig. \ref{fig:magnetization}. 

\section{On the behavior of the susceptibility}

Another important function of physical interest is based on the two-point correlation functions 
\begin{equation}
	\label{eq:two-point-correlation-function}
	g_{ij}(N,K,K_{ia},h)=\la S_i S_j \ra -\la S_i\ra \la S_j \ra.
\end{equation}
With the help of $\left\{g_{ij}\right\}$, i,j=1,$\cdots, N$ one has the the susceptibility is
\begin{equation}
	\label{eq:susceptibility}
	\chi(N,K,K_{ia},h)=\frac{1}{N} \sum_{i=1}^N \sum_{j=1}^N g_{ij} (N,K,K_{ia},h). 
\end{equation} 
It is straightforward to see that
\begin{eqnarray}
	\label{eq:susceptibility-relations}
	\chi=\frac{\dd m}{\dd h} &=&  N \left[\bigg \la \left(\frac{1}{N}\sum_{i=1}^N S_i\right)^2 \bigg \ra_{\mu_N} - \left(\bigg\la \frac{1}{N}\sum_{i=1}^N S_i \bigg\ra_{\mu_N} \right)^2\right] \nonumber \\
	&=& N \bigg \la \left[\frac{1}{N}\sum_{i=1}^N S_i- \bigg\la \frac{1}{N}\sum_{i=1}^N S_i \bigg\ra_{\mu_N}  \right]^2 \bigg \ra_{\mu_N}.
\end{eqnarray}
From \eq{eq:susceptibility-relations} it follows that $\chi(N,K,K_{ia},h)>0$. Having in mind \eq{eq:average-magnetization}, we derive that
\begin{equation}
	\label{eq:susceptibility-free-energy}
	\chi(N,K,K_{ia},h)=-\frac{\dd^2}{\dd h^2} \left[\beta f(N,K,K_{ia},h)\right]>0,
\end{equation}
and, therefore, that $m$ is an odd increasing function of $h$.
\begin{figure}[htb]
	\centering
	\includegraphics[scale=.46]{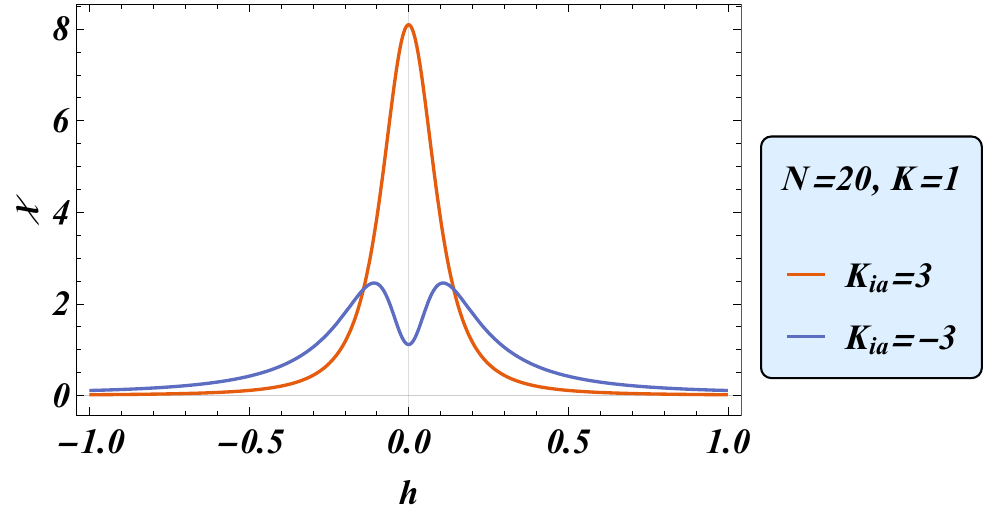} \quad \includegraphics[scale=.46]{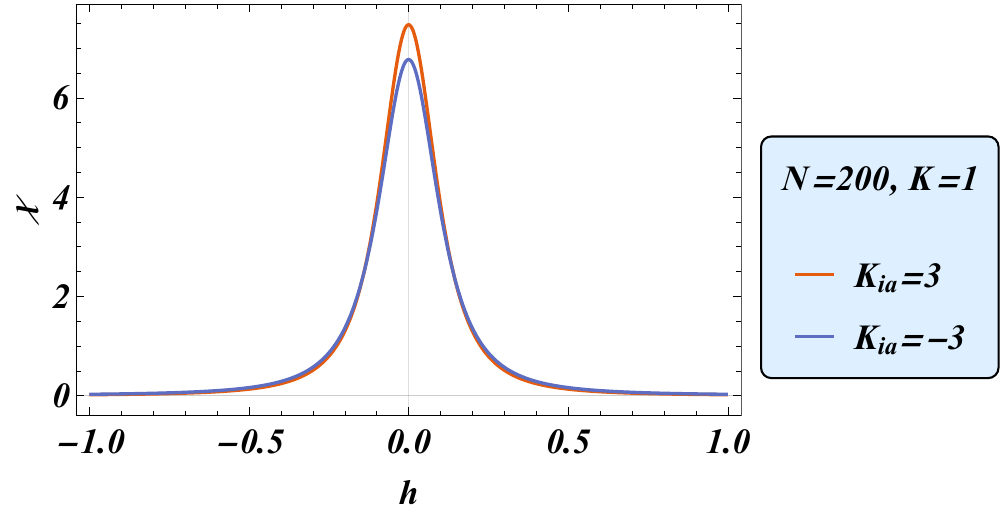} 
	\caption{As we see, the susceptibility is symmetric with respect of the sign of the external magnetic field $h$. In the current figure the \textbf{left panel} shows that for relatively small $N$ one has a single minimum in a chain in which the impurity atom interacts in a way similar to the rest of the chain (i.e., in a ferromagnetic way; $K_{ia=3}$), while in the opposite case, $K_{ia}=-3$, it has two maxima.  The \textbf{right panel}: for relatively large $N$ the both functions are very similar. }
	\label{fig:susceptibility}     
\end{figure}
The behavior of the susceptibility as a function of $h$ for $N=20$ and $N=200$, with $K=1$ and  $K_{ia}=\pm 3$ is visualized in Fig. \ref{fig:susceptibility}. 

\section{Behavior of the Casimir Force in Ising Chain with Impurity Atom}
\label{sec:Casimir-force-plots}

In accordance with the definitions Eqs. \eqref{CasDef}, \eqref{excess_free_energy_definition}, from 
Eqs. 	\eqref{eq:decomposition-of-free-energy} -- 	\eqref{eq:remnant-part} for the Casimir force we obtain the following result
\begin{equation}
	\label{eq:Casimir-nonzero-field}
	\beta F_{\rm Cas}(N,K,K_{ia},h)=-\frac{1}{N} \frac{N}{ \xi (K,h)} \frac{1}{r^{-1}(K,K_{ia},h) \exp \left[(N-2)/\xi (K,h)\right]+1}<0.
\end{equation}
Obviously, the sigh of the force is determined by the sign of $r^{-1}(K,K_{ia},h)$; since  $r^{-1}(K,K_{ia},h)>0$ the force is \textit{attractive}.

\begin{figure}[htb]
	\centering
	\includegraphics[scale=.5]{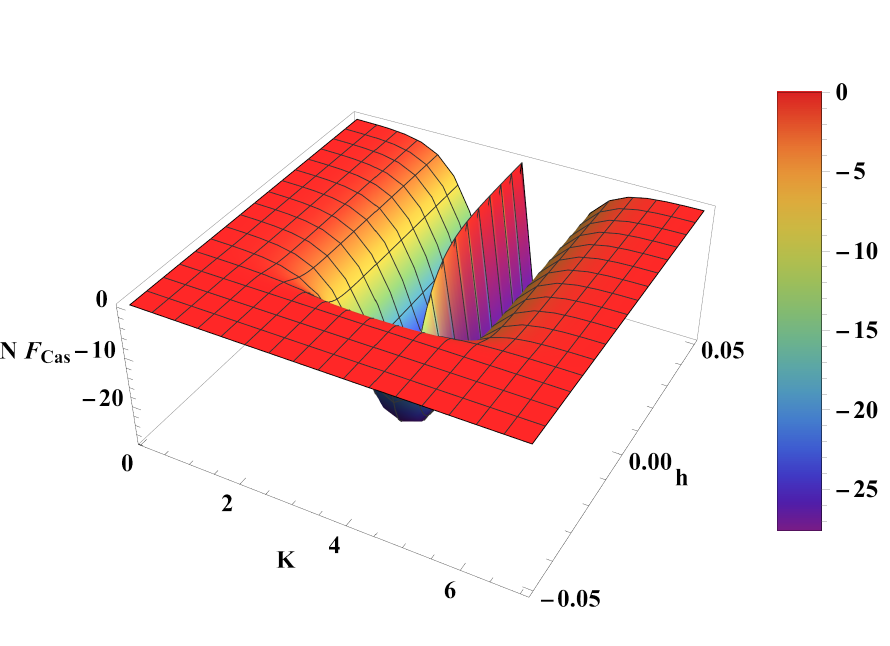} \quad \includegraphics[scale=.5]{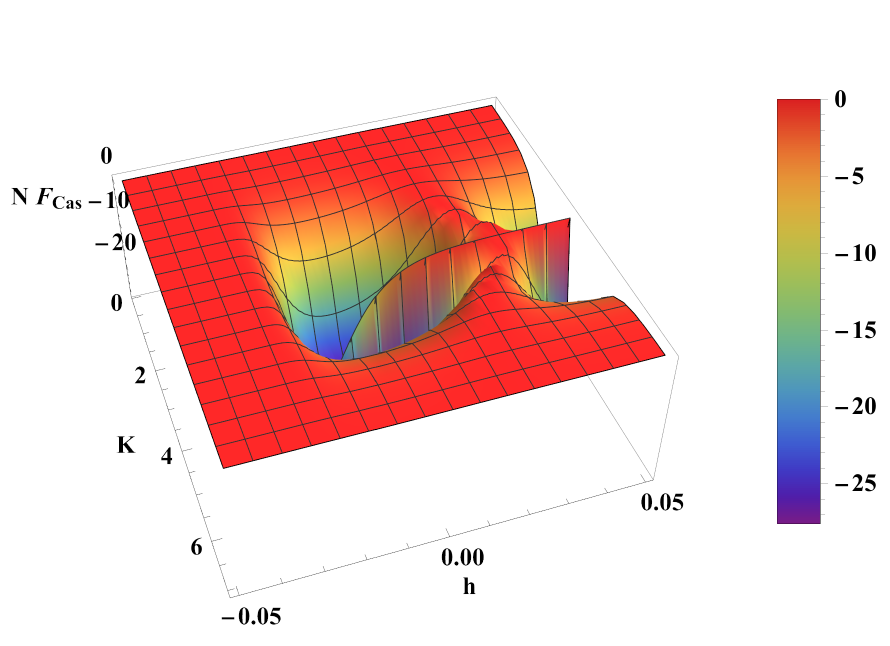} 
	\caption{As we see, the force is always symmetric with respect of the sign of the external magnetic field $h$. In the current figure the \textbf{left panel} shows the behavior of the Casimir force for $N=100$ with $K_{ia}=3$. The \textbf{right panel} depicts  the behavior of the Casimir for again  $N=100$ but for $K_{ia}=-3$. The behavior of the force is shown as a function of $K$ and $h$.}
	\label{fig:Casimir-3D}     
\end{figure}

The behavior of the Casimir force as a function of $K$, $K_{ia}$ and $h$ for $N=100$ and $K_{ia}=\pm 3$ is shown in Fig. 	\ref{fig:Casimir-3D}.     

Let us now consider the behavior of the force in the scaling regime. On general grounds we expect that the scaling function of the Casimir force $X_{\rm Cas}(x_t,x_a,x_h)$ is 
\begin{equation}
	\label{eq:scaling-function-Casimir-definition}
	\beta F_{\rm Cas}(N,K,K_{ia}h)=\frac{1}{N} X_{\rm Cas}(x_t,K_{ia},x_h).
\end{equation}
From \eq{eq:scaling_variables_Ising}, and Eqs. \eqref{eq:Casimir-nonzero-field} -- \eqref{eq:scaling-function-Casimir-definition} one derives the corresponding explicit expressions 
\begin{equation}
	\label{eq:scaling-function-Casimir}
	X_{\rm Cas}(K,K_{ia},h)=-\frac{\sqrt{x_h^2+x_t^2}}{r^{-1}(x_t,K_{ia},x_h) \exp \left(\sqrt{x_h^2+x_t^2}\right) +1}, 
\end{equation}
where
\begin{equation}
	\label{eq:scaling-Casimir}
	r(K,K_{ia},h)= \left[\frac{\left(1-2 e^{2 (K_{ia}-K)}\right) }{(2 \cosh K_{ia})^2}\left(\frac{x_h}{x_t}\right)^2+\tanh (K_{ia})\right]^2.
\end{equation}
Obviously, since  $r(K,K_{ia},h)>0$ one has $X_{\rm Cas}(x_t,K_{ia},x_h)<0$. Furthermore,  $X_{\rm Cas}(x_t,K_{ia},x_h)$ decays exponentially when $x_h^2+x_t^2\gg 1$. It is clear, however, that perfect scaling exist only for $K_{ia}=K$.

section{Concluding Comments and Discussion}
\label{sec:conclusion}

In the current article we have studied the behavior of the one-dimensional Ising model of Ising chain with impurity atom. We determined analytically the partition function and the free energy of the system. Then, we found specific results for the behavior of the magnetization (i.e., of the average spin), see Fig. \ref{fig:magnetization}, of the response function, i.e., of the susceptibility -- see Fig. \ref{fig:susceptibility}, and of the Casimir fluctuation induced fore -- see Fig. \ref{fig:Casimir-3D}. Let us note that the behavior of the one-dimensional infinite Ising chain can be found in many books on statistical mechanics - see, e.g., \cite{MW73,H87,B82,Kardar2007,Pathria2011}. The behavior of the finite chain has been studied in  \cite{Rudnick2010,DD2022,Dantchev2023,Dantchev2024,tonchev2024chebyshev} where periodic, antiperiodic and Dirichlet boundary conditions have been imposed. Here we consider the special case of Ising chain in the form of a ring in which one impurity atom is present. Note that, contrary to the case of a single defect bond, where the Casimir force can be both positive and negative, depending on the sign and the strength of the coupling constant characterizing this bond, in the situation of an impurity atom this is not the case - the force is always attractive.

\begin{acknowledgments}
	Partial financial support via Grant No KP-06-H72/5 of the Bulgarian NSF is gratefully acknowledged.
\end{acknowledgments}

\nocite{*}

\begin{thebibliography}{15}%
	\makeatletter
	\providecommand \@ifxundefined [1]{%
		\@ifx{#1\undefined}
	}%
	\providecommand \@ifnum [1]{%
		\ifnum #1\expandafter \@firstoftwo
		\else \expandafter \@secondoftwo
		\fi
	}%
	\providecommand \@ifx [1]{%
		\ifx #1\expandafter \@firstoftwo
		\else \expandafter \@secondoftwo
		\fi
	}%
	\providecommand \natexlab [1]{#1}%
	\providecommand \enquote  [1]{``#1''}%
	\providecommand \bibnamefont  [1]{#1}%
	\providecommand \bibfnamefont [1]{#1}%
	\providecommand \citenamefont [1]{#1}%
	\providecommand \href@noop [0]{\@secondoftwo}%
	\providecommand \href [0]{\begingroup \@sanitize@url \@href}%
	\providecommand \@href[1]{\@@startlink{#1}\@@href}%
	\providecommand \@@href[1]{\endgroup#1\@@endlink}%
	\providecommand \@sanitize@url [0]{\catcode `\\12\catcode `\$12\catcode
		`\&12\catcode `\#12\catcode `\^12\catcode `\_12\catcode `\%12\relax}%
	\providecommand \@@startlink[1]{}%
	\providecommand \@@endlink[0]{}%
	\providecommand \url  [0]{\begingroup\@sanitize@url \@url }%
	\providecommand \@url [1]{\endgroup\@href {#1}{\urlprefix }}%
	\providecommand \urlprefix  [0]{URL }%
	\providecommand \Eprint [0]{\href }%
	\providecommand \doibase [0]{http://dx.doi.org/}%
	\providecommand \selectlanguage [0]{\@gobble}%
	\providecommand \bibinfo  [0]{\@secondoftwo}%
	\providecommand \bibfield  [0]{\@secondoftwo}%
	\providecommand \translation [1]{[#1]}%
	\providecommand \BibitemOpen [0]{}%
	\providecommand \bibitemStop [0]{}%
	\providecommand \bibitemNoStop [0]{.\EOS\space}%
	\providecommand \EOS [0]{\spacefactor3000\relax}%
	\providecommand \BibitemShut  [1]{\csname bibitem#1\endcsname}%
	\let\auto@bib@innerbib\@empty
	\bibitem [{\citenamefont {Krech}(1994)}]{Krech1994}%
	\BibitemOpen
	\bibfield  {author} {\bibinfo {author} {\bibfnamefont {M.}~\bibnamefont
			{Krech}},\ }\href@noop {} {\emph {\bibinfo {title} {The Casimir Effect in
				Critical Systems}}}\ (\bibinfo  {publisher} {World Scientific, Singapore},\
	\bibinfo {year} {1994})\BibitemShut {NoStop}%
	\bibitem [{\citenamefont {Brankov}, \citenamefont {Dantchev},\ and\
		\citenamefont {Tonchev}(2000)}]{BDT2000}%
	\BibitemOpen
	\bibfield  {author} {\bibinfo {author} {\bibfnamefont {J.~G.}\ \bibnamefont
			{Brankov}}, \bibinfo {author} {\bibfnamefont {D.~M.}\ \bibnamefont
			{Dantchev}}, \ and\ \bibinfo {author} {\bibfnamefont {N.~S.}\ \bibnamefont
			{Tonchev}},\ }\href@noop {} {\emph {\bibinfo {title} {The Theory of Critical
				Phenomena in Finite-Size Systems - Scaling and Quantum Effects}}}\ (\bibinfo
	{publisher} {World Scientific, Singapore},\ \bibinfo {year}
	{2000})\BibitemShut {NoStop}%
	\bibitem [{\citenamefont {Macio\l{}ek}\ and\ \citenamefont
		{Dietrich}(2018)}]{MD2018}%
	\BibitemOpen
	\bibfield  {author} {\bibinfo {author} {\bibfnamefont {A.}~\bibnamefont
			{Macio\l{}ek}}\ and\ \bibinfo {author} {\bibfnamefont {S.}~\bibnamefont
			{Dietrich}},\ }\bibfield  {title} {\enquote {\bibinfo {title} {Collective
				behavior of colloids due to critical {C}asimir interactions},}\ }\href
	{\doibase 10.1103/RevModPhys.90.045001} {\bibfield  {journal} {\bibinfo
			{journal} {Rev. Mod. Phys.}\ }\textbf {\bibinfo {volume} {90}},\ \bibinfo
		{pages} {045001} (\bibinfo {year} {2018})}\BibitemShut {NoStop}%
	\bibitem [{\citenamefont {Dantchev}\ and\ \citenamefont
		{Dietrich}(2023)}]{DD2022}%
	\BibitemOpen
	\bibfield  {author} {\bibinfo {author} {\bibfnamefont {D.~M.}\ \bibnamefont
			{Dantchev}}\ and\ \bibinfo {author} {\bibfnamefont {S.}~\bibnamefont
			{Dietrich}},\ }\bibfield  {title} {\enquote {\bibinfo {title} {Critical
				casimir effect: Exact results},}\ }\href {\doibase
		https://doi.org/10.1016/j.physrep.2022.12.004} {\bibfield  {journal}
		{\bibinfo  {journal} {Phys. Rep.}\ }\textbf {\bibinfo {volume} {1005}},\
		\bibinfo {pages} {1--130} (\bibinfo {year} {2023})}\BibitemShut {NoStop}%
	\bibitem [{\citenamefont {Gambassi}\ and\ \citenamefont
		{Dietrich}(2024)}]{Gambassi2024}%
	\BibitemOpen
	\bibfield  {author} {\bibinfo {author} {\bibfnamefont {A.}~\bibnamefont
			{Gambassi}}\ and\ \bibinfo {author} {\bibfnamefont {S.}~\bibnamefont
			{Dietrich}},\ }\bibfield  {title} {\enquote {\bibinfo {title} {Critical
				casimir forces in soft matter},}\ }\href {\doibase 10.1039/d3sm01408h}
	{\bibfield  {journal} {\bibinfo  {journal} {Soft Matter}\ } (\bibinfo {year}
		{2024}),\ 10.1039/d3sm01408h}\BibitemShut {NoStop}%
	\bibitem [{\citenamefont {McCoy}\ and\ \citenamefont {Wu}(1973)}]{MW73}%
	\BibitemOpen
	\bibfield  {author} {\bibinfo {author} {\bibfnamefont {B.~M.}\ \bibnamefont
			{McCoy}}\ and\ \bibinfo {author} {\bibfnamefont {T.~T.}\ \bibnamefont {Wu}},\
	}\href@noop {} {\emph {\bibinfo {title} {The Two-dimensional Ising Model}}}\
	(\bibinfo  {publisher} {Harvard Univ. Press, Cambridge},\ \bibinfo {year}
	{1973})\BibitemShut {NoStop}%
	\bibitem [{\citenamefont {Huang}(1987)}]{H87}%
	\BibitemOpen
	\bibfield  {author} {\bibinfo {author} {\bibfnamefont {K.}~\bibnamefont
			{Huang}},\ }\href@noop {} {\emph {\bibinfo {title} {Statistical
				Mechanics}}},\ \bibinfo {edition} {2nd}\ ed.\ (\bibinfo  {publisher} {John
		Wiley \& Sons, New York},\ \bibinfo {year} {1987})\ p.\ \bibinfo {pages}
	{512}\BibitemShut {NoStop}%
	\bibitem [{\citenamefont {Baxter}(1982)}]{B82}%
	\BibitemOpen
	\bibfield  {author} {\bibinfo {author} {\bibfnamefont {R.~J.}\ \bibnamefont
			{Baxter}},\ }\href@noop {} {\emph {\bibinfo {title} {Exactly Solved Models in
				Statistical Mechanics}}}\ (\bibinfo  {publisher} {Academic, London},\
	\bibinfo {year} {1982})\BibitemShut {NoStop}%
	\bibitem [{\citenamefont {Kardar}(2007)}]{Kardar2007}%
	\BibitemOpen
	\bibfield  {author} {\bibinfo {author} {\bibfnamefont {M.}~\bibnamefont
			{Kardar}},\ }\href@noop {} {\emph {\bibinfo {title} {Statistical Physics of
				Fields}}}\ (\bibinfo  {publisher} {Cambridge University Press, Cambridge},\
	\bibinfo {year} {2007})\BibitemShut {NoStop}%
	\bibitem [{\citenamefont {Pathria}\ and\ \citenamefont
		{Beale}(2011)}]{Pathria2011}%
	\BibitemOpen
	\bibfield  {author} {\bibinfo {author} {\bibfnamefont {R.~K.}\ \bibnamefont
			{Pathria}}\ and\ \bibinfo {author} {\bibfnamefont {P.~D.}\ \bibnamefont
			{Beale}},\ }\href@noop {} {\emph {\bibinfo {title} {Statistical
				Mechanics}}},\ \bibinfo {edition} {3rd}\ ed.\ (\bibinfo  {publisher}
	{Elsevier, Amsterdam},\ \bibinfo {year} {2011})\BibitemShut {NoStop}%
	\bibitem [{\citenamefont {Rudnick}\ \emph {et~al.}(2010)\citenamefont
		{Rudnick}, \citenamefont {Zandi}, \citenamefont {Shackell},\ and\
		\citenamefont {Abraham}}]{Rudnick2010}%
	\BibitemOpen
	\bibfield  {author} {\bibinfo {author} {\bibfnamefont {J.}~\bibnamefont
			{Rudnick}}, \bibinfo {author} {\bibfnamefont {R.}~\bibnamefont {Zandi}},
		\bibinfo {author} {\bibfnamefont {A.}~\bibnamefont {Shackell}}, \ and\
		\bibinfo {author} {\bibfnamefont {D.}~\bibnamefont {Abraham}},\ }\bibfield
	{title} {\enquote {\bibinfo {title} {Boundary conditions and the critical
				{C}asimir force on an {I}sing model film: Exact results in one and two
				dimensions},}\ }\href {\doibase 10.1103/PhysRevE.82.041118} {\bibfield
		{journal} {\bibinfo  {journal} {Phys.Rev.E}\ }\textbf {\bibinfo {volume}
			{82}},\ \bibinfo {pages} {041118} (\bibinfo {year} {2010})}\BibitemShut
	{NoStop}%
	\bibitem [{\citenamefont {Dantchev}, \citenamefont {Tonchev},\ and\
		\citenamefont {Rudnick}(2023)}]{Dantchev2023}%
	\BibitemOpen
	\bibfield  {author} {\bibinfo {author} {\bibfnamefont {D.~M.}\ \bibnamefont
			{Dantchev}}, \bibinfo {author} {\bibfnamefont {N.~S.}\ \bibnamefont
			{Tonchev}}, \ and\ \bibinfo {author} {\bibfnamefont {J.}~\bibnamefont
			{Rudnick}},\ }\bibfield  {title} {\enquote {\bibinfo {title} {Casimir versus
				{H}elmholtz forces: Exact results},}\ }\href {\doibase
		https://doi.org/10.1016/j.aop.2023.169533} {\bibfield  {journal} {\bibinfo
			{journal} {Annals of Physics}\ }\textbf {\bibinfo {volume} {459}},\ \bibinfo
		{pages} {169533} (\bibinfo {year} {2023})}\BibitemShut {NoStop}%
	\bibitem [{\citenamefont {Dantchev}, \citenamefont {Tonchev},\ and\
		\citenamefont {Rudnick}(2024)}]{Dantchev2024}%
	\BibitemOpen
	\bibfield  {author} {\bibinfo {author} {\bibfnamefont {D.}~\bibnamefont
			{Dantchev}}, \bibinfo {author} {\bibfnamefont {N.}~\bibnamefont {Tonchev}}, \
		and\ \bibinfo {author} {\bibfnamefont {J.}~\bibnamefont {Rudnick}},\
	}\bibfield  {title} {\enquote {\bibinfo {title} {Casimir and {H}elmholtz
				forces in one-dimensional {I}sing model with {D}irichlet (free) boundary
				conditions},}\ }\href {\doibase https://doi.org/10.1016/j.aop.2024.169647}
	{\bibfield  {journal} {\bibinfo  {journal} {Annals of Physics}\ }\textbf
		{\bibinfo {volume} {464}},\ \bibinfo {pages} {169647} (\bibinfo {year}
		{2024})}\BibitemShut {NoStop}%
	\bibitem [{\citenamefont {Tonchev}\ and\ \citenamefont
		{Dantchev}(2024)}]{tonchev2024chebyshev}%
	\BibitemOpen
	\bibfield  {author} {\bibinfo {author} {\bibfnamefont {N.~S.}\ \bibnamefont
			{Tonchev}}\ and\ \bibinfo {author} {\bibfnamefont {D.}~\bibnamefont
			{Dantchev}},\ }\bibfield  {title} {\enquote {\bibinfo {title} {Chebyshev
				polynomials in the physics of the one-dimensional finite-size {I}sing model:
				An alternative view and some new results},}\ }\href@noop {} {\bibfield
		{journal} {\bibinfo  {journal} {Condensed Matter}\ }\textbf {\bibinfo
			{volume} {9}},\ \bibinfo {pages} {53} (\bibinfo {year} {2024})}\BibitemShut
	{NoStop}%
	\bibitem [{\citenamefont {Liebchen}\ and\ \citenamefont
		{Löwen}(2019)}]{Liebchen2019}%
	\BibitemOpen
	\bibfield  {author} {\bibinfo {author} {\bibfnamefont {B.}~\bibnamefont
			{Liebchen}}\ and\ \bibinfo {author} {\bibfnamefont {H.}~\bibnamefont
			{Löwen}},\ }\bibfield  {title} {\enquote {\bibinfo {title} {Which
				interactions dominate in active colloids?}}\ }\href {\doibase
		10.1063/1.5082284} {\bibfield  {journal} {\bibinfo  {journal} {The Journal of
				Chemical Physics}\ }\textbf {\bibinfo {volume} {150}},\ \bibinfo {pages}
		{061102} (\bibinfo {year} {2019})},\ \Eprint
	{http://arxiv.org/abs/https://doi.org/10.1063/1.5082284}
	{https://doi.org/10.1063/1.5082284} \BibitemShut {NoStop}%
\end{thebibliography}

%


\end{document}